\documentclass[conference]{IEEEtran}
\IEEEoverridecommandlockouts
\usepackage{tabularx}
\usepackage[T1]{fontenc}% optional T1 font encoding
\usepackage{algorithm}
\usepackage[noend]{algpseudocode}
\usepackage{graphicx}
\usepackage{amsmath,amssymb,amsfonts,stfloats}
\usepackage{array}
\usepackage{textcomp}
\usepackage[font=small]{caption}
\usepackage{paralist}
\usepackage{lettrine}
\usepackage{cite}
\usepackage{xcolor}
\usepackage{cancel}
\usepackage{booktabs}
\usepackage{mathtools}
\usepackage{array}
\usepackage{booktabs}
\usepackage[caption=false,font=footnotesize]{subfig}
\usepackage{url}
\usepackage{acronym}
\usepackage{multicol}
\usepackage{makecell}

\begin{document}

\title{Network-Controlled Repeaters vs. Reconfigurable Intelligent Surfaces for 6G mmW Coverage Extension}
\author{Reza Aghazadeh Ayoubi,
        Marouan Mizmizi, Dario Tagliaferri, Danilo De Donno and Umberto Spagnolini}

%\markboth{Journal of \LaTeX\ Class Files,~Vol.~14, No.~8, February~2022}%
%{Shell \MakeLowercase{\textit{et al.}}: Bare Demo of IEEEtran.cls for IEEE Journals}

\maketitle

\begin{abstract}
Network-controlled repeaters (NCR) and reconfigurable intelligent surfaces (RIS) are being considered by the third generation partnership project (3GPP) as valid candidates for range extension in millimeter-wave (mmW, 30-300 GHz frequency) 5G and 6G networks, to counteract large path and penetration losses. Nowadays, there is no definite answer on which of the two technologies is better for coverage extension, since existing comparative studies focused either on communication-related metrics only or on very large-scale analysis, without physical layer details or environmental considerations.  
This paper aims at comparing NCR and RIS solutions in terms of coverage in an urban scenario. Numerical results suggest a possible usage of both technologies, showing that preferring RIS or NCR depends not only on their physical capabilities and cost (e.g., the number of antenna elements, the angular separation between panels for NCR, etc.) but also on environment geometry.

%Comparison of the reflective intelligent surface (RIS) with classic relays is a crucial need for optimum network planning and deployment, which has been analysed in several works. Previous works have either focused on such comparison from the point of view of communication without considering scenario and environment-dependant parameters such as geometry and blockage loss, or they have focused on large-scale network analysis without physical layer subtleties such as realistic channel models and antenna element directivity patterns.

%This paper performs a comparative coverage analysis of RIS versus the steering-enabled amplify and forward (AF) relays, namely smart repeaters (NCR). 
%The numerical results suggest possible hybrid application of these two network entities in a large scale wireless network, by showing that the preference between the RIS and SR is governed not only on the capabilities of the RIS and SR and their deployment cost and power consumption in network level, but also on limitations imposed by: i) interference mitigation measures; ii) antenna element directivity; iii) environment geometry and blockage; and iv) the positions of the network entities.
\end{abstract}

\IEEEpeerreviewmaketitle
\section{Introduction}
Future 6G communication systems are envisaged to massively exploit the millimeter waves (mmW - $30-300$ GHz) frequencies to enable capacity-eager verticals, e.g., autonomous driving and extended reality \cite{6G_beyond}. The propagation at mmW suffers a severe path and penetration loss, which limits the network coverage and link reliability. Multiple-input multiple-output (MIMO) systems using large antenna arrays are, \textit{de-facto}, the present solution to overcome the current challenges \cite{6G_Vision}. Still, the coverage and capacity performance in harsh or dense urban environments are not satisfactory \cite{BJORNSON20193}, calling for cost-effective, low-power, and sustainable technology to improve network performance.

Among the investigated technologies, reconfigurable intelligent surface (RIS) and network-controlled repeater (NCR) are the most promising. RIS is a novel system node leveraging smart radio surfaces with thousands of sub-wavelength metamaterial elements to dynamically shape and control radio signals in a goal-oriented manner \cite{Renzo2020,CIRS}. NCR consists of two analog arrays of antennas, one oriented towards the base station (BS) and the other towards the user equipments. NCR is the evolution of the wireless repeater based on amplify-and-forward (AF) concept, with advanced capabilities, such as beamforming and time-division duplexing (TDD) operations to support uplink and downlink modes \cite{HuaweiSR, Qualcomm}. While academia and industry are thoroughly exploring technical problems and use cases of these technologies, early standardization processes kick-off to further drive the progress of RIS and NCR. Indeed, the third-generation partnership project (3GPP) has launched a study item on NCR for its Release 18 \cite{3GPP_WG1}, while the European telecommunications standards institute (ETSI) started an industrial Specification Group    on RIS \cite{RIS_ETSI}.

NCRs use a mature technology that has been consolidated over the years, while RISs have been proposed only recently, and their full potential is yet to be unlocked. Nevertheless, understanding which technologies and under what conditions might be used, are necessary to accelerate market introduction. The authors in \cite{DF_vs_RIS} compare a half and full duplex decode-and-forward (DF) relay and a RIS. They show that the RIS only surpasses the half-duplex scheme in some specific cases, while an ideal full-duplex DF scheme always surpasses the RIS. In \cite{DiRenzo_RIS_vs_DF}, the RIS configured as a focusing lens might overtake the DF performance, although this requires exact knowledge of the position of the Tx and Rx. Hybrid RIS-relay schemes have also been proposed in \cite{Coexistence_CoDesign, RelayRisCooperation}, indicating that a higher data rate can be achieved considering both solutions. 
The authors in \cite{Eugenio1} carry out a large-scale mmW network planning analysis, showing that given a constant budget and costs of network radio entities, the existence of RIS substantially increases the coverage and throughput of the access network. In \cite{RIS_RS_Eugenio}, the co-existence of RIS and NCR in mmW radio access network is analyzed, showing that given different parameters, joint installation of the RISs and NCRs is preferred.

This paper provides a comparative analysis of RIS and NCR from the physical layer point of view in a realistic urban environment, where static and dynamic blockages are considered in the channel model. The static blockage model is deterministic and based on the digital map, while dynamic blockage is modeled based on 3GPP recommendations. Extensive numerical simulations show that both solutions can provide coverage extension and help mitigate link blockage. Moreover, the results show that the NCR's limits are the angular separation between antenna arrays and the served field of view (FoV). High angular separation reduces the flexibility of deployment and, subsequently, the coverage, but it is necessary to guarantee tolerable levels of self-interference. Therefore, NCR is indicated for those environments where all the UEs can be served within a limited FoV (e.g., open roads). On the other side, RIS exhibits higher coverage capabilities compared to NCR, although with less capacity, at the price of a non-negligible number of elements posing practical challenges related to configurability \cite{Renzo2020}, which are not addressed in this paper.

The remainder of the paper is organized as follows: Section \ref{sect:systemmodel} details the system and channel model in presence of RIS and NCR, Section \ref{sect:blockage} outlines the employed blockage model, while Section \ref{sect:results} reports the numerical results. Finally, Section \ref{sect:conclusion} concludes the paper.

The adopted notation is as follows: bold upper- and lower-case letters describe matrices and column vectors. Matrix transposition, conjugation, conjugate transposition and Frobenius norm are indicated respectively as $\mathbf{A}^{\mathrm{T}}$, $\mathbf{A}^{*}$, $\mathbf{A}^{\mathrm{H}}$ and $\|\mathbf{A}\|_F$. $\mathrm{diag}(\mathbf{a})$ is the diagonal matrix given by vector $\mathbf{a}$. $\mathbf{I}_n$ is the identity matrix of size $n$. $|\mathcal{A}|$ denotes the cardinality of set $\mathcal{A}$. With  $\mathbf{a}\sim\mathcal{CN}(\boldsymbol{\mu},\mathbf{C})$ we denote a multi-variate circularly complex Gaussian random variable. $\mathbb{E}[\cdot]$ is the expectation operator, while $\mathbb{R}$ and $\mathbb{C}$ stand for the set of real and complex numbers, respectively. $\delta_{n}$ is the Kronecker delta.
%
%In this paper, we provide a comparative analysis from the physical layer point of view using the realistic channel model of the RIS and NCR and static and dynamic blockage models. The static blockage model is deterministic and based on the digital map of a portion of the city of Milan. NCR is chosen here for comparison specifically because: i) NCR is a newer and promising technology w.r.t the classical AF and DF schemes \cite{SRE,HuaweiSR}; ii) \gls{DF} relays are practically expensive to deploy and introduce large processing delays; iii) beam steering capability of NCR is one of the key measures that contributes to the mitigation of loop-back interference in full duplex relays \cite{Emil_RIS_DF_Interference}. We show numerically that the angular separation between the NCR panels imposed by the residual self-interference suppression measures, antenna element directivity pattern and panels shielding, substantially affects the preferred choice between the RIS and NCR, based on the areas that each entity can serve better. 
%
%We provide an exemplary scenario to show how different parameters of the radio entities and also the geometry of the environment, affect the performance of each entity. We demonstrate that one of the crucial aspects that has usually not been considered or has been neglected are the antenna elements' directivity and the angular distance of the panels of repeaters. This study paves the way to perform a more accurate analysis for large-scale network planning optimization similar to \cite{RIS_RS_Eugenio}.
%
\section{System Model}\label{sect:systemmodel}
\begin{figure}[b!]
\centering
\includegraphics[width=0.9\columnwidth]{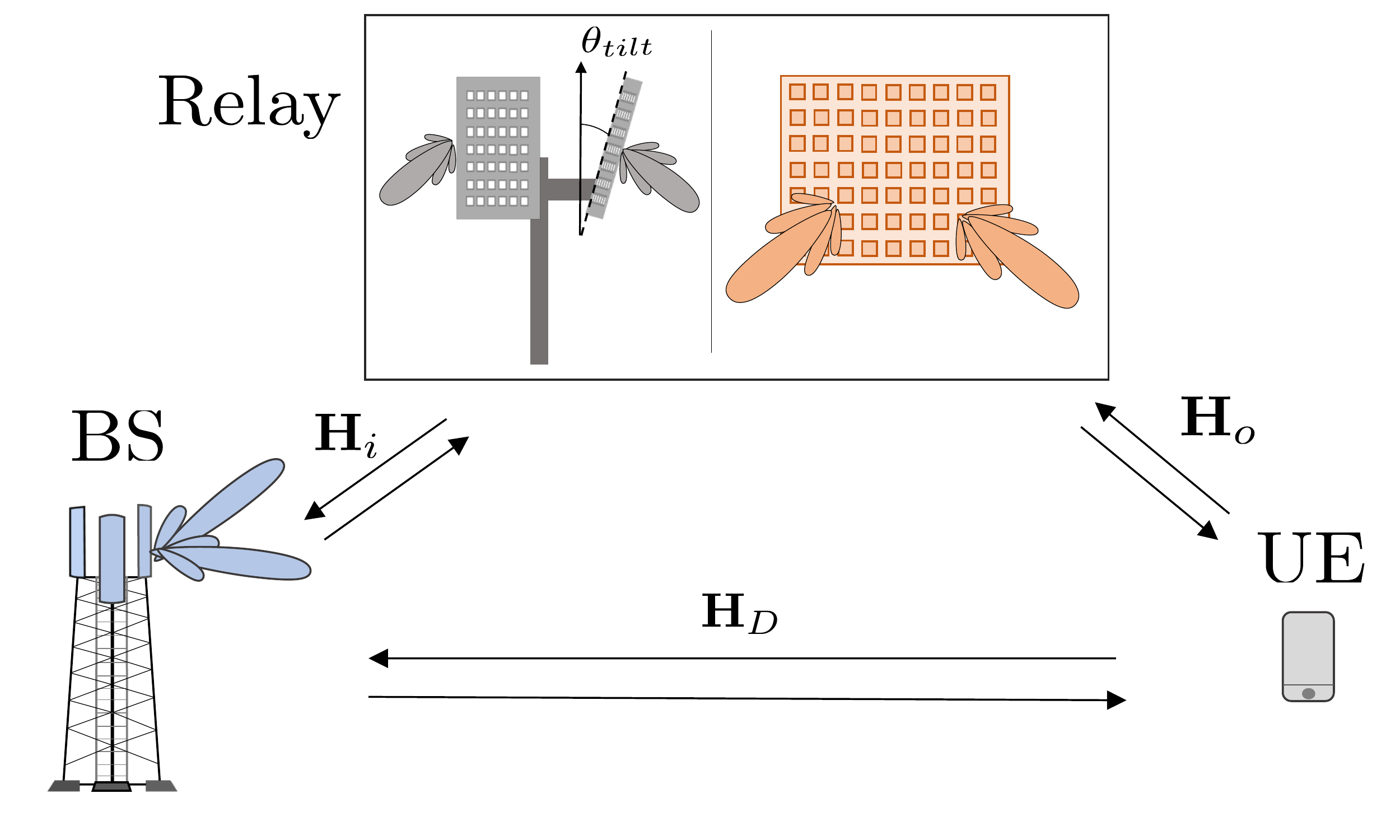}
\caption{\small The scheme of direct and relay-aided communication} \label{fig:SysModel}
\end{figure}

Let us consider the downlink communication scenario depicted in Fig. \ref{fig:SysModel}, with two possible links (direct and/or relayed) between a Tx (BS) and a single Rx (UE). The BS, relay and UE positions in space, defined in a global coordinate system, are denoted as $\mathbf{p}_{BS} = [x_{BS}, y_{BS}, z_{BS}]^\mathrm{T}$, $\mathbf{p}_{R} = [x_{R}, y_{R}, z_{R}]^\mathrm{T}$, $\mathbf{p}_{UE} = [x_{UE}, y_{UE}, z_{UE}]^\mathrm{T}$ respectively. The set of all the possible UE positions is indicated with $\mathcal{P}=\{\mathbf{p}_{UE}\}$. BS and UE are equipped with antenna arrays of $N_t$ and $N_r$ elements, respectively. 
The relay can be either a RIS or a two-panel AF NCR. The RIS has $M$ elements, while the NCR has two panels with $N_{p}$ elements each. The first panel of NCR is pointed toward the BS while the second panel is oriented toward the coverage area. All the considered arrays are half-wavelength-spaced at center bandwidth, i.e., around carrier frequency $f_0=c/\lambda$ (for speed of light $c$ and $\lambda$ wavelength), while the elements of the RIS are $\lambda/4$ spaced.

\subsection{Signal Model}\label{subsect:signal_model}

Let the vector of complex Tx symbols be $\mathbf{s} \in \mathbb{C}^{N_s\times 1} \sim \mathcal{CN}\left(0, (\sigma_s^2/N_s)\mathbf{I}_{N_s}\right)$, where $N_s$ is the number of data streams and $\sigma_s^2$ is the total Tx power. Symbols $\mathbf{s}$ are precoded by matrix $\mathbf{F} \in \mathbb{C}^{N_t \times N_s}$, obtaining the Tx signal
\begin{equation}
    \mathbf{x} = \mathbf{F}\,\mathbf{s},
\end{equation}
such that $\mathrm{tr}(\mathbf{F}\mathbf{F}^\mathrm{H})=N_t$. The propagation is over a block-faded spatially-sparse channel $\mathbf{H} \in \mathbb{C}^{N_r \times N_t}$, whose model depends on whether we are considering the RIS or the NCR. In case of RIS, after the time-frequency synchronization, the Rx signal $\mathbf{y}\in \mathbb{C}^{N_s \times 1}$ is
\begin{equation}\label{eq:receivedSignal_RIS}
    \mathbf{y}_{ris} = \mathbf{W}^\mathrm{H} (\mathbf{H}_d + \mathbf{H}_{ris}) \mathbf{F} \,\mathbf{s} + \mathbf{W}^\mathrm{H} \mathbf{n},
\end{equation}
where $\mathbf{W} \in \mathbb{C}^{N_r \times N_s}$ is the Rx combiner at UE, such that $\mathrm{tr}(\mathbf{W}\mathbf{W}^\mathrm{H})=N_r$, $\mathbf{H}_d \in\mathbb{C}^{N_r \times N_t}$ is the direct BS-UE channel, $\mathbf{H}_{ris} \in\mathbb{C}^{N_r \times N_t}$ is the BS-RIS-UE channel and the additive noise is $\mathbf{n} \sim \mathcal{CN}(\mathbf{0}, \sigma^2_n\mathbf{I}_{N_r})$.
In the case of an NCR, the model must include the noise amplified at the relay. Hence, we have
\begin{equation}\label{eq:receivedSignal_SR}
    \mathbf{y}_{ncr} = \mathbf{W}^\mathrm{H} (\mathbf{H}_d + \mathbf{H}_{ncr}) \mathbf{F} \,\mathbf{s} + \mathbf{W}^\mathrm{H} \mathbf{H}_{z} \mathbf{z} + \mathbf{W}^\mathrm{H} \mathbf{n},
\end{equation}
where $\mathbf{H}_{ncr} \in\mathbb{C}^{N_r \times N_t}$ is the BS-NCR-UE channel, $\mathbf{z} \sim \mathcal{CN}(\mathbf{0}, \sigma^2_z\mathbf{I}_{N_p})$ is the noise generated at the NCR Rx antennas and $\mathbf{H}_{z}\in\mathbb{C}^{N_r \times N_p}$ is the propagation channel for the noise $\mathbf{z}$. The channel models are presented in the following. 

\subsection{Channel model}\label{subsect:channel_model in presence of RIS}

\subsubsection{Direct Channel} The model for the direct channel $\mathbf{H}_d$ can be written for the far-field assumption, that typically applies for Tx-Rx distances in the order of tens of meters, as \cite{rappaport2019wireless}:
\begin{equation}\label{eq:channelModel}
    \mathbf{H}_d = \sum_{p=1}^{P} \alpha_p\, \varrho_r(\boldsymbol{\vartheta}_p^r) \varrho_t(\boldsymbol{\vartheta}_p^t) \mathbf{a}_r(\boldsymbol{\vartheta}_p^r)\mathbf{a}_t(\boldsymbol{\vartheta}_p^t)^\mathrm{H}, 
\end{equation}
where \textit{(i)} $P$ is the number of paths \textit{(ii)} $\alpha_p$ denotes the scattering amplitude of the $p$-th path, \textit{(ii)} $\varrho_t(\boldsymbol{\vartheta}_p^t)$ and $\varrho_t(\boldsymbol{\vartheta}_p^r)$ are the Tx and Rx single-antenna gains for the $p$-th path, respectively, defined according to \cite{3GPP} as functions of the Tx and Rx pointing angles $\boldsymbol{\vartheta}_p^t$ and $\boldsymbol{\vartheta}_p^r$, \textit{(iii)} $\mathbf{a}_t(\boldsymbol{\vartheta}_p^t)\in\mathbb{C}^{N_t\times 1}$ and $\mathbf{a}_r(\boldsymbol{\vartheta}_p^r)\in\mathbb{C}^{N_r\times 1}$ are the Tx and Rx array response vectors for the $p$-th path.   

\subsubsection{Channel through the RIS} The channel through the RIS is customarily modelled as \cite{tang2020wireless}:
\begin{equation}\label{eq:channel_RIS}
  \mathbf{H}_{ris} = \mathbf{H}_{o} \boldsymbol{\Phi} \mathbf{H}_{i}.    
\end{equation}
Specifically, $\mathbf{H}_{i} \in \mathbb{C}^{M \times N_t}$ is the BS-RIS MIMO channel for the incident signal, $\mathbf{H}_{o} \in \mathbb{C}^{N_r \times M}$ is the RIS-UE MIMO channel for the reflected signal, and $\boldsymbol{\Phi}\in \mathbb{C}^{M \times M}$ is the phase configuration matrix of the RIS. 
%\begin{figure}[t!]
%\begin{centering}
%\includegraphics[scale=0.40]{Content/Figures/RIS.pdf}
%\end{centering}
%\caption{\textcolor{red}{dow we need this?} \label{fig:RIS_Rot} Elevation and azimuth of the incoming wave to the RIS with index 'i', and outgoing waves from the RIS with index 'o'}
%\end{figure}
While the far-field assumption generally holds for the direct BS-UE paths, the link through the RIS can be either in far-field (with planar wavefront) or near-field (with curved wavefront) depending on the size of the RIS compared to the BS/UE-RIS distances. Therefore, the $(m,n)$-th entries of $\mathbf{H}_{i}$ and $(m,k)$-th entries $\mathbf{H}_{o}$ are
\begin{align}\label{eq:generalModel}
    &\left[\mathbf{H}_{i}\right]_{m,n} = \alpha_{m,n}\; \varrho_t(\boldsymbol{\vartheta}_{m,n}^t) \, \varrho_{ris}(\boldsymbol{\vartheta}_{m,n}^{i}) \, e^{-j\frac{2\pi}{\lambda}r_{m,n}^t},\\
    &\left[\mathbf{H}_{o}\right]_{m,k} = \beta_{m,k}\;
    \varrho_{ris}(\boldsymbol{\vartheta}_{m,k}^{o})
    \varrho_r(\boldsymbol{\vartheta}_{m,k}^r) \, e^{-j\frac{2\pi}{\lambda}r_{m,k}^r},
\end{align}
where \textit{(i)} $\alpha_{m,n}$ and $\beta_{m,k}$ are the complex gains of the path between the $n$-th BS antenna and the $m$-th RIS element and between the $m$-th RIS element and the $k$-th UE antenna, respectively \cite{Gustafson6691924, tang2020wireless}; \textit{(ii)} $r_{m,n}^t$ and $r_{m,k}^r$ are propagation distances; \textit{(iii)} $\varrho_{ris}(\cdot)$, is the single element pattern of the RIS, function of incidence and reflection angles $\boldsymbol{\vartheta}_{m,n}^{i}$ and $\boldsymbol{\vartheta}_{m,k}^{o}$, based on the widely employed cosine model for reflectarrays \cite{EllingsonPathLoss, RadPatt}. The reflection matrix $\boldsymbol{\Phi}$ in \eqref{eq:channel_RIS} is diagonal with entries defined as
\begin{equation}\label{eq:reflectingmatrix}
    \boldsymbol{\Phi} = \text{diag}\left( e^{j\Phi_{1}},...,e^{j\Phi_{m}},..., e^{j\Phi_{M}}\right),
\end{equation}
where $\Phi_{m}$ denotes the phase applied at the $m$-th element. The overall channel impulse response between the $n$-th BS array element and the $k$-th UE array element is
\begin{equation}\label{eq:focusingChannel_singlaPath}
    \left[\mathbf{H}_{ris}\right]_{(n,k)} = 
    \sum_{m} \Gamma_{m,n,k} 
    e^{-j\left(\frac{2\pi}{\lambda}(r_{m,k}^t + r_{m,n}^r) + \Phi_m\right)},
\end{equation}
where $\Gamma_{m,n,k}$ incorporates the path loss and the element's gain in the RIS channel. 

\begin{figure}[t!]
\centering
\includegraphics[width=0.8\columnwidth]{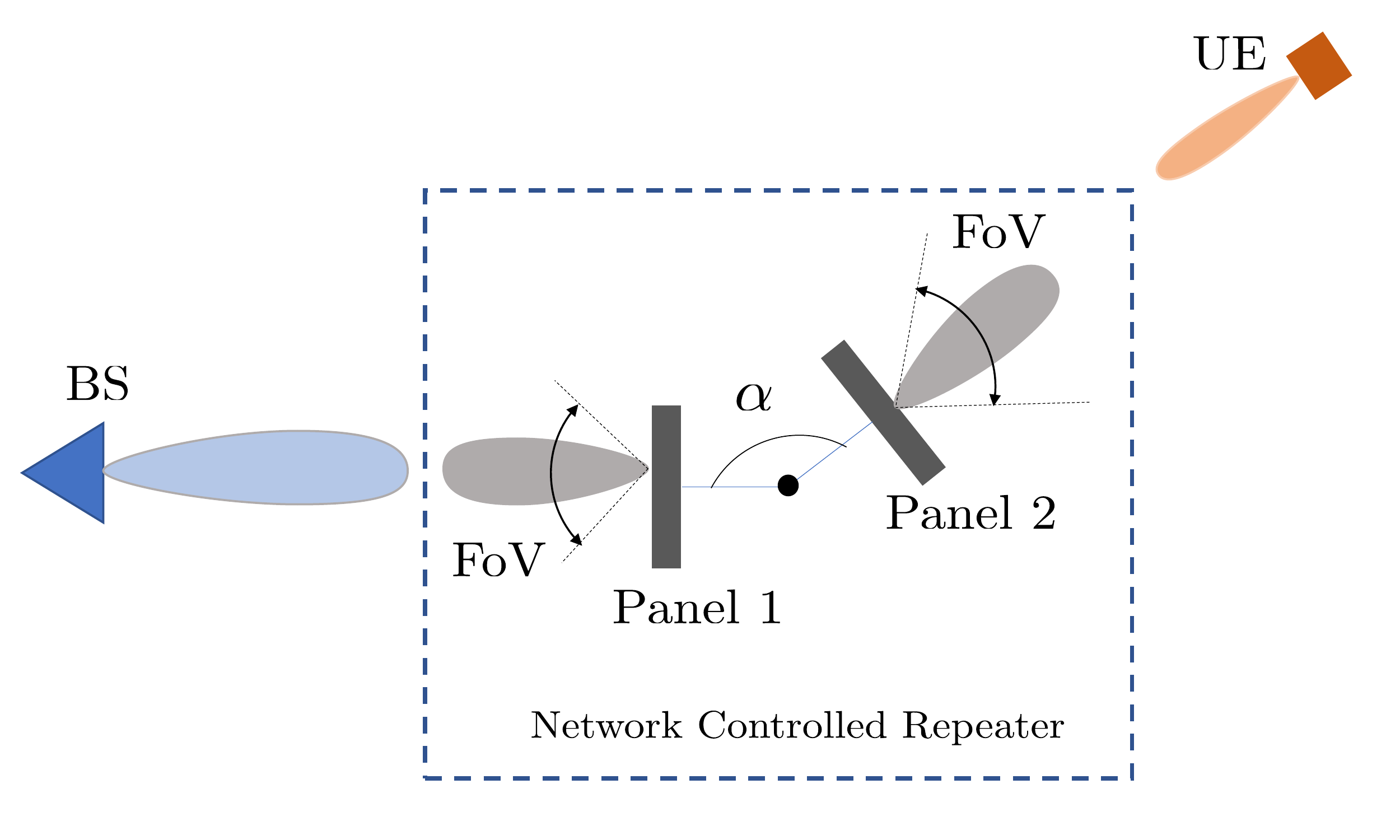}
  \caption{\small Map view of a network-controlled repeater (NCR). $\alpha$ is the angular separation of the two panels}
  \label{fig:SR_scheme}
\end{figure}

\subsubsection{Channel through the NCR} %Differently from the RIS which are fully-passive, 
The NCR is an active relay capable of amplification. The scheme of the NCR is shown in Fig.~\ref{fig:SR_scheme}, where an angular separation $\alpha$ between the two panels is enforced to limit the loop-back interference (e.g., $\alpha = 120$ deg \cite{HuaweiSR}). The channel model is \cite{3SectSR}: 
\begin{equation}\label{eq:channel_NCR}
    \mathbf{H}_{ncr} = g\,\mathbf{H}_{o}\mathbf{f}_{p}\mathbf{w}_{p}^{H}\mathbf{H}_{i},
\end{equation}
where \textit{(i)} $g$ is the amplification factor,% provided by the NCR, 
\textit{(ii)} $\mathbf{H}_{i} \in \mathbb{C}^{N_{p} \times N_t}$ and $\mathbf{H}_{o} \in \mathbb{C}^{N_r \times N_{p}}$ are the BS-NCR and NCR-UE    channels, respectively, \textit{(iii)} $\mathbf{w}_{p}\in \mathbb{C}^{N_{p} \times 1}$ is the analog combiner at the NCR for the impinging signal from the BS, such that $\|\mathbf{w}_{p}\|_2 = N_p$ and \textit{(iv)}  $\mathbf{f}_{p}\in \mathbb{C}^{N_{p}\times 1}$ is the analog precoder at the NCR panel toward the UE, such that $\|\mathbf{f}_{p}\|_2 = N_p$. Channels $\mathbf{H}_{i}$ and $\mathbf{H}_{o}$ are modelled as in \eqref{eq:channelModel}, with straightforward modifications.
The channel for noise $\mathbf{z}$ in \eqref{eq:receivedSignal_SR} is 
\begin{equation}
    \mathbf{H}_z = g\,\mathbf{H}_{o}\mathbf{f}_{p}\mathbf{w}_{p}^{H}.
\end{equation}
The E2E power gain $G$ of the NCR is
\begin{equation}
    G = |g|^2 \mathrm{tr}\,(\mathbf{f}_{p}\mathbf{w}_{p}^{\mathrm{H}}\mathbf{w}_{p}\mathbf{f}_{p}^{\mathrm{H}}) = |g|^2 N_p^2,
\end{equation}
that takes into account the array and element gains (of the cascade of the two panels) and the amplifier gain. 
The amplification gain $g$ must be tuned such that $G<G_{max}$, where $G_{max}$ is provided by the NCR specifications from 3GPP.

%\textcolor{red}{do we need this? can't we simply refer to our works?} \subsubsection{Path-Loss and Fading}

%The path amplitudes in \eqref{eq:channelModel} and \eqref{eq:generalModel} depend on the path loss $PL$:
%
%\begin{equation}\label{eq:complexGain}
%    \alpha = \sqrt{\frac{1}{PL}} e^{j\xi}
%\end{equation}
%
%where the phase $\xi$ accounts for additional effects (e.g., Doppler shift) and it is assumed as uniformly distributed, i.e., $\xi\sim\mathcal{U}[0, 2\pi)$, independent across different paths. The propagation loss for the direct link with amplitude $\alpha_d$ is defined (in dB) \cite{3GPPTR37885} as
%
%\begin{equation} \label{eq:PL0}
%   PL_d = \mu_{LoS} + A_b + \chi
%\end{equation}
%
%where $\mu_{LoS}$ depends on the Tx-Rx distance and on the scenario considered, $\chi \sim \mathcal{N}(0, \sigma_{sh}^2)$ represents the log-normal distributed shadowing component, and $A_b \sim \mathcal{N}(\mu_b, \sigma^2_b)$ accounts for an additional attenuation due to blockage from $b$ vehicles simultaneously \cite{R1-1807672, 3GPPTR37885}.

%Assuming independence between the shadowing component and the blockage component, we can write the path-loss as
%\begin{equation}\label{eq:tot_pl}
%    PL_d \sim
    %\mathcal{N}\left(\underbrace{\mu_{\text{LoS}}+\mu_b}_{\mu_{PL_b}}, \underbrace{\sigma_{sh}^2 + \sigma^2_b}_{\sigma_{PL_b}^2}\right)
%\end{equation}
%

\section{Blockage model}\label{sect:blockage}
%\subsection{Blockage probability}

The amplitudes of the channels in \eqref{eq:channelModel} are modelled following the mmW path-loss model proposed in \cite{3GPPTR37885}, comprising the free-space loss, the log-normal shadowing and an additional attenuation for blockage. The blockage is either static or dynamic. The static blockage is environment-specific and can be \textit{a-priori} evaluated using the buildings' geometry extracted from digital maps as well as the locations of the BS, NCR/RIS and UEs. 
Differently, the dynamic blockage due to mobile obstacles (e.g., cars, trucks, etc.) at mmW can only be modelled with stochastic approaches. For the purpose of this work, we need to model the blockage probability of a given mmW link of length $r$. Based on the work in \cite{DynamicBlockage}, the blockage probability can be modelled as
\begin{equation}\label{eq:blockage_probabilty}
    \mathrm{P}_B(r) = \frac{r \mu^{-1}C  }{1 + r\mu^{-1} C },
\end{equation}
where $\mu^{-1}$ is the blockage duration and $C$ is defined as
\begin{equation}
    C = \frac{2}{\pi} \lambda_B V \frac{z_B - z_T}{z_T - z_R},
\end{equation}
in which $\lambda_B$ and $V$ are the dynamic blockers' average spatial density and velocity, while $z_T$, $z_B$ and $z_R$ are the transmitters, blocker and receiver heights, respectively.  
%\subsection{Blockage loss}
The dynamic blockage loss due to human bodies and vehicles is based on the 3GPP model B, modelling the loss in dB scale as \cite{3GPPTR38901}
\begin{equation}
    L_{\mathrm{B}}=-20\log_{10}{\left(1-\left(F_{h_1}+F_{h_2}\right)\left(F_{w_1}+F_{w_2}\right)\right)},
\end{equation}
where $F_{h_1}$, $F_{h_2}$, $F_{w_1}$ and $F_{w_2}$ account for knife edge diffraction coefficients defined in \cite{3GPPTR38901}.

\begin{figure*}[t!]
\vspace{4mm}
\centering
\subfloat[\small Direct links]{\includegraphics[width=0.30\textwidth]{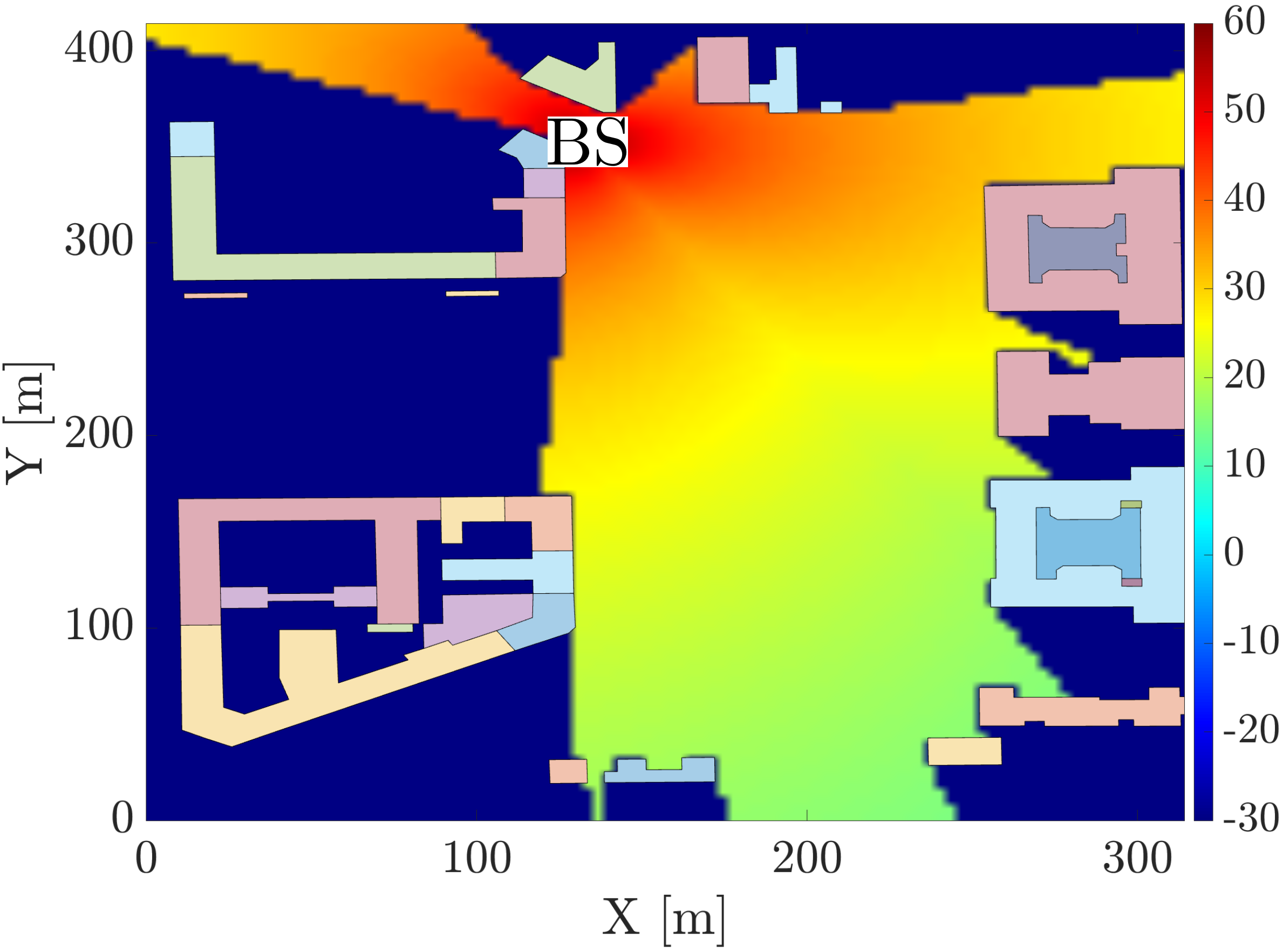}
  \label{fig:Dir} }
  \subfloat[\small RIS only links]{ \includegraphics[width=0.30\textwidth]{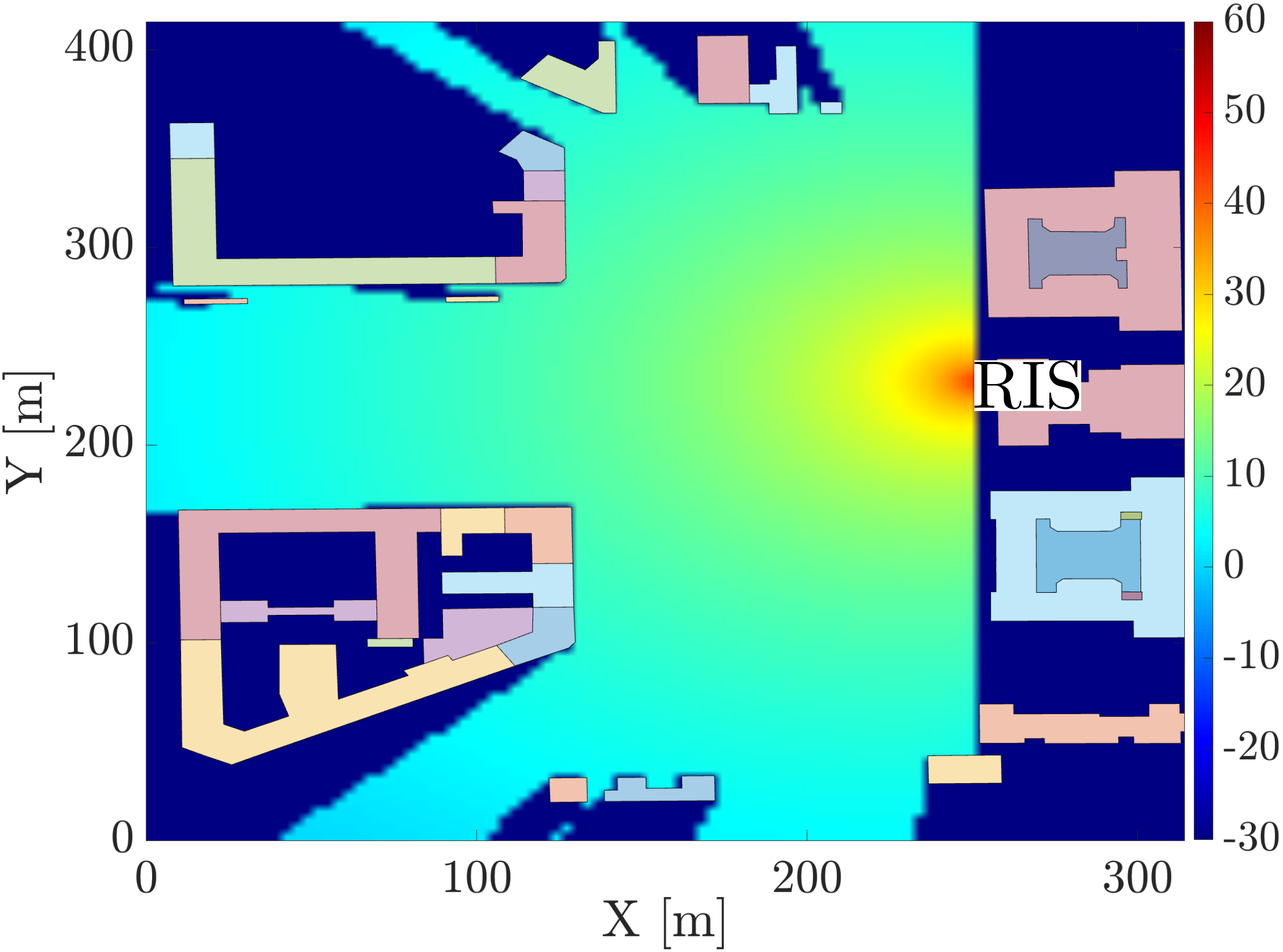}
  \label{fig:RIS} }
  \subfloat[\small NCR only links]{\includegraphics[width=0.30\textwidth]{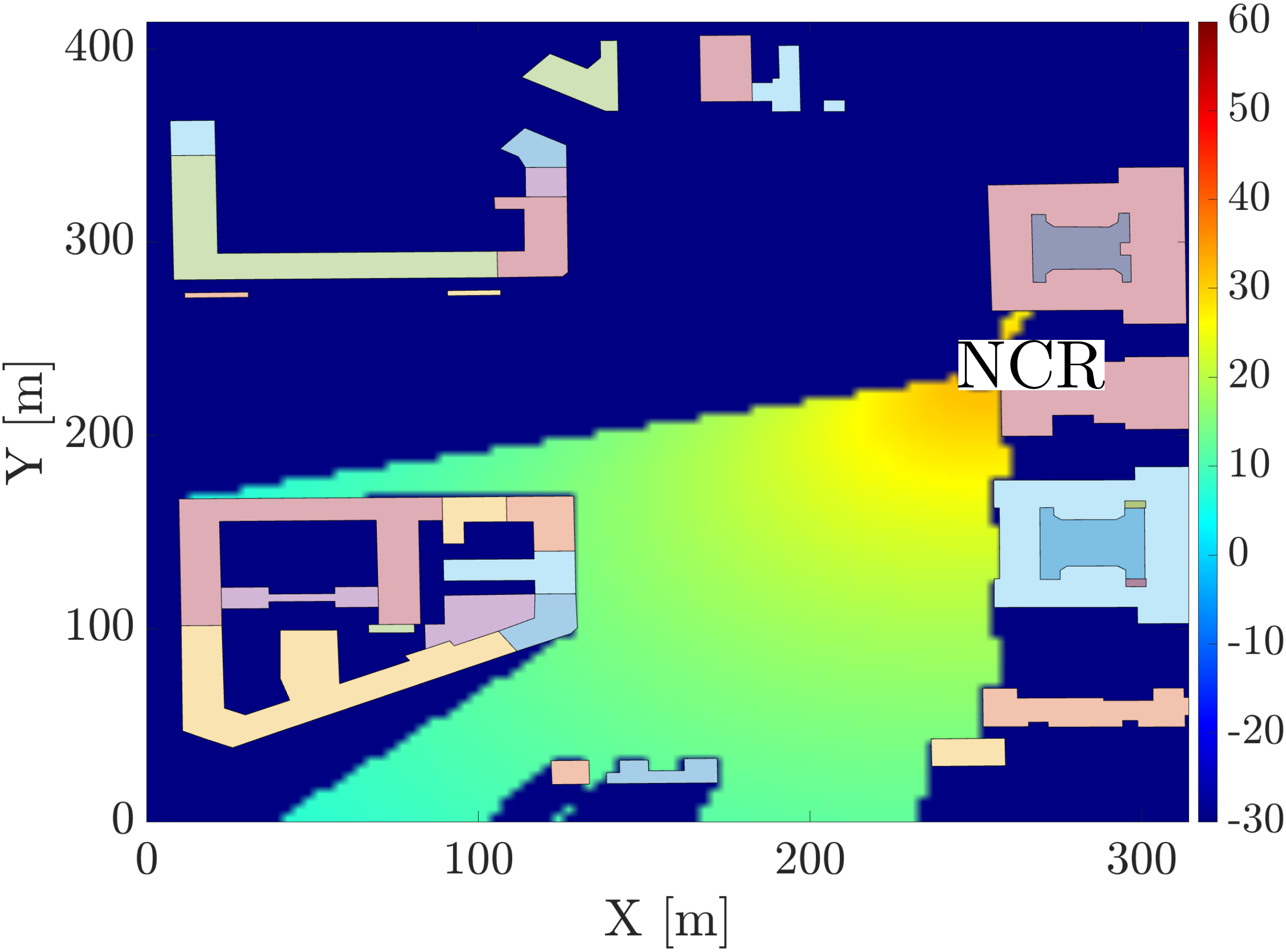}
  \label{fig:SR} }
  \caption{\small Heamap of the SNR perceived by a generic UE at position $\mathbf{p}_{UE} \in \mathcal{P}$, in urban scenario 1, served by: a)  BS through direct path; b) RIS as relay; c) NCR as relay. The positions of the BS and NCR/RIS are fixed.  \label{fig:Heatmap}}
\end{figure*}

\begin{figure*}[t!]
\centering
\subfloat[\small Direct Links]{\includegraphics[width=0.31\textwidth]{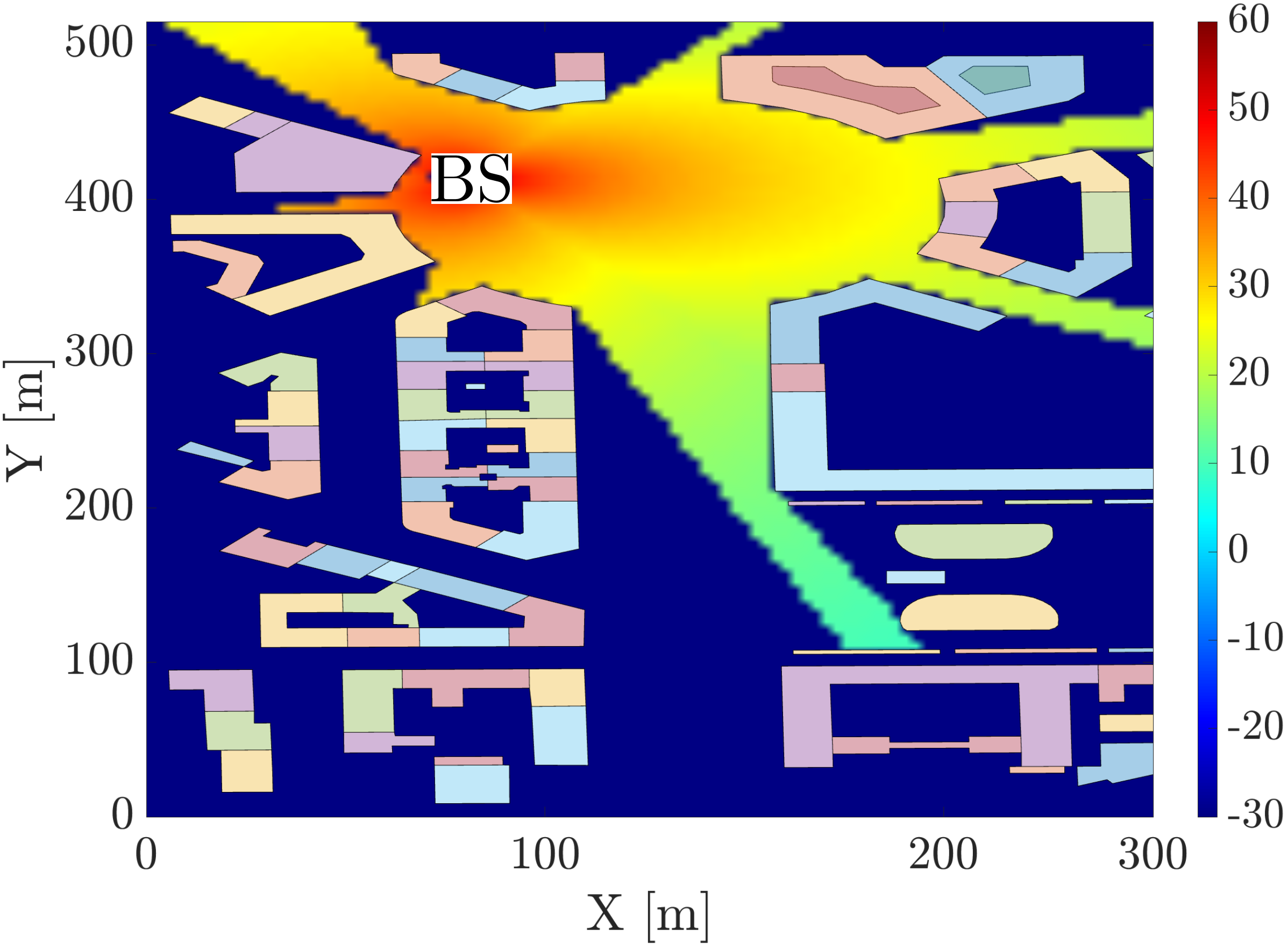}
  \label{subfig:Dir2} }
  \subfloat[\small RIS only links]{ \includegraphics[width=0.31\textwidth]{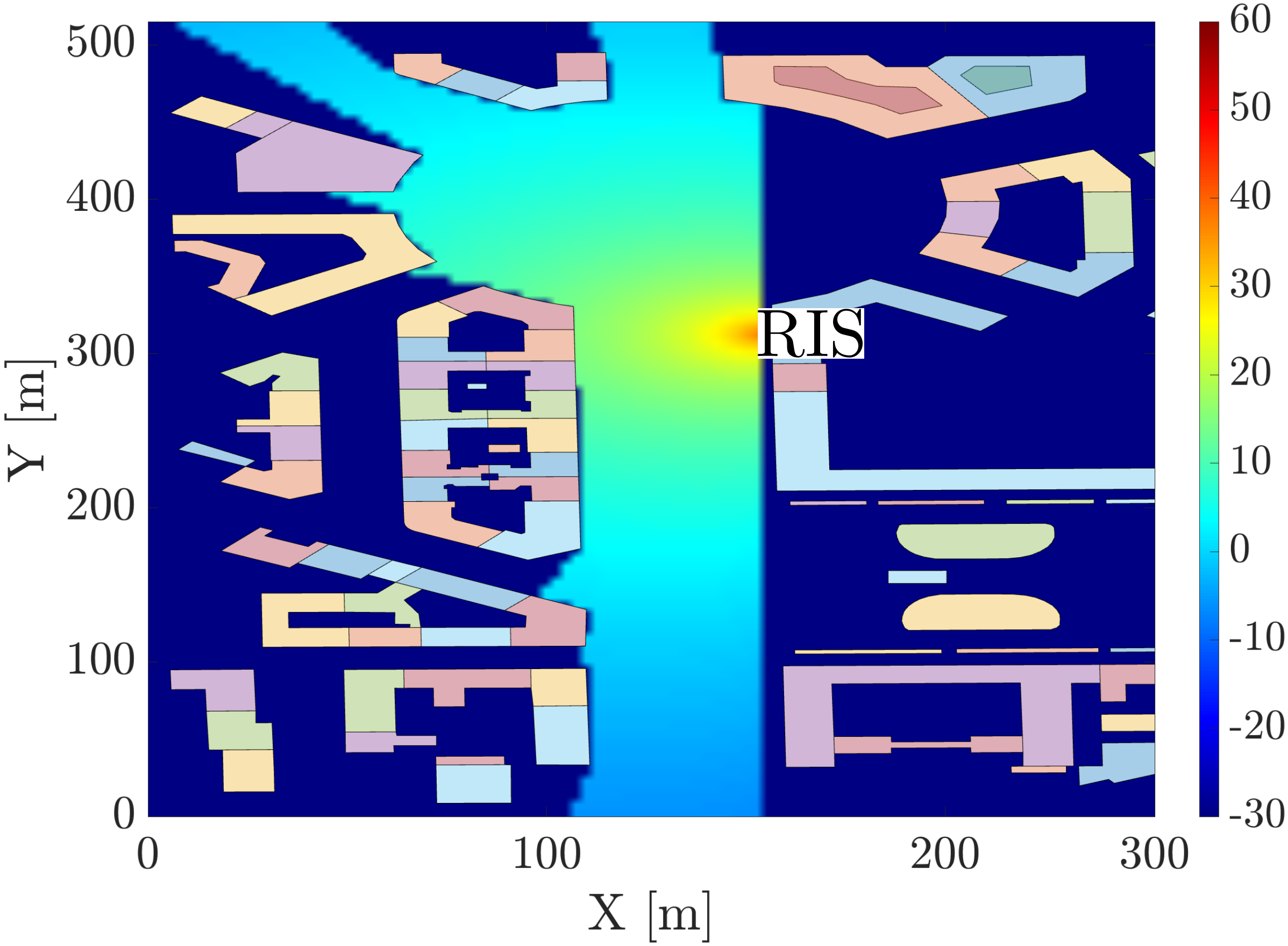}
  \label{subfig:RIS2} }
  \subfloat[\small NCR only links]{\includegraphics[width=0.31\textwidth]{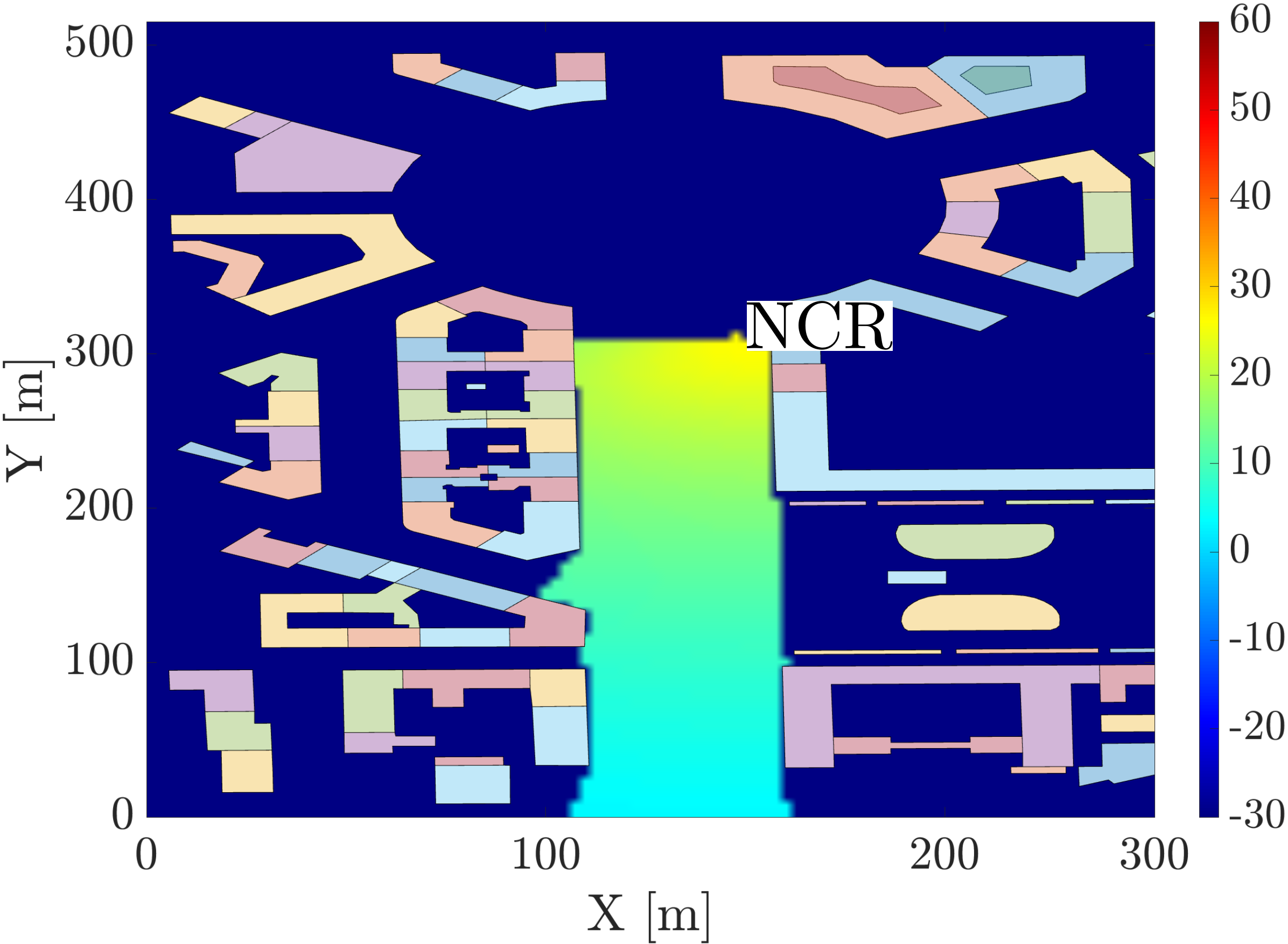}
  \label{subfig:SR2} }
  \caption{ \small Heamap of the SNR perceived by a generic UE at position $\mathbf{p}_{UE} \in \mathcal{P}$, in urban scenario 2, served by: a)  BS through direct path; b) RIS as relay; c) NCR as relay. The positions of the BS and NCR/RIS are fixed.  \label{fig:Heatmap2}}
\end{figure*}

\begin{figure*}
\centering
\subfloat[][\small RIS-aided, urban scenario 1]{\includegraphics[width=0.42\textwidth]{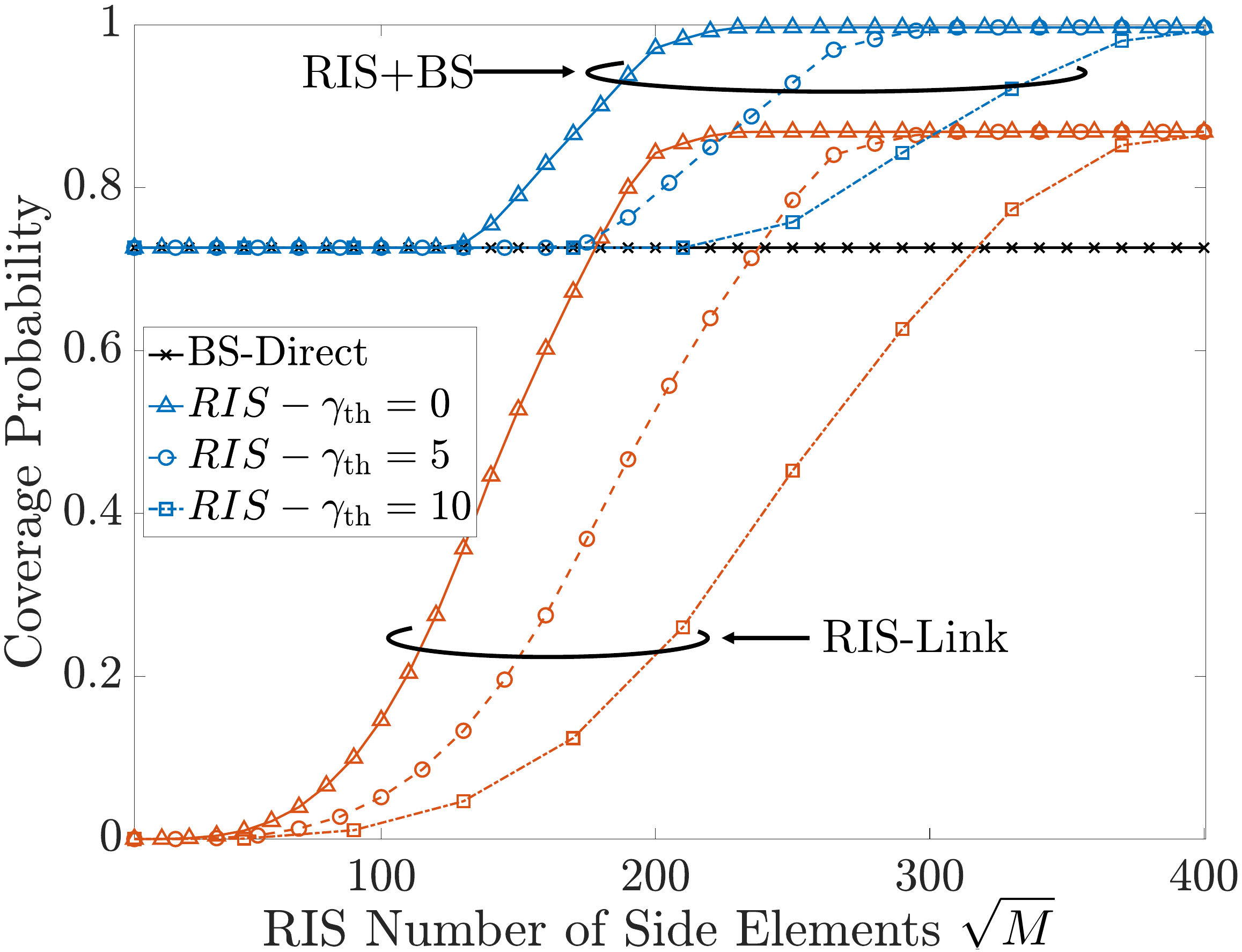}\label{fig:SC1_RIS}} \hspace{0.5cm}
\subfloat[][\small NCR-aided, urban scenario 1]{\includegraphics[width=0.42\textwidth]{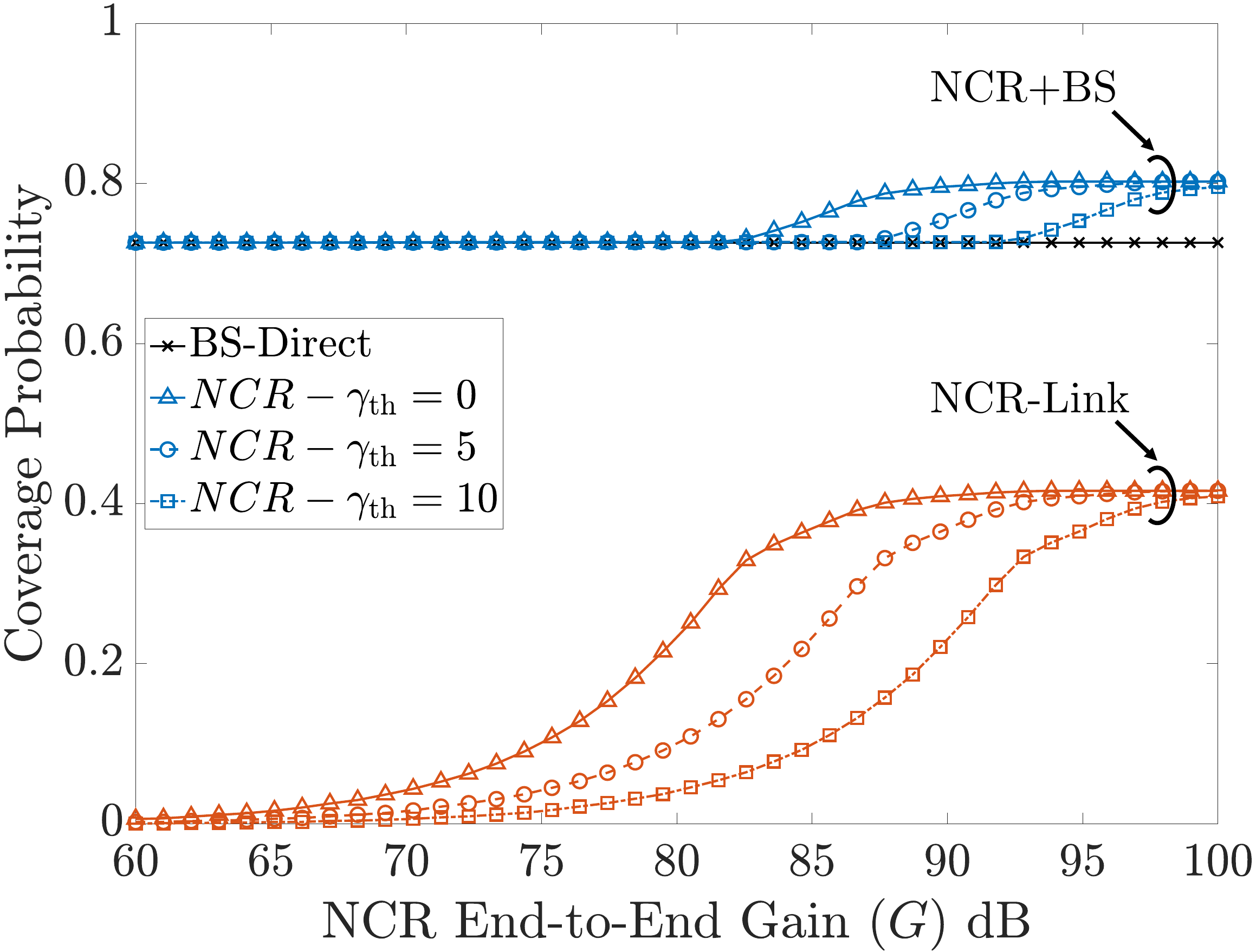}\label{fig:SC1_SR}}\\
\subfloat[][\small RIS-aided, urban scenario 2]{\includegraphics[width=0.42\textwidth]{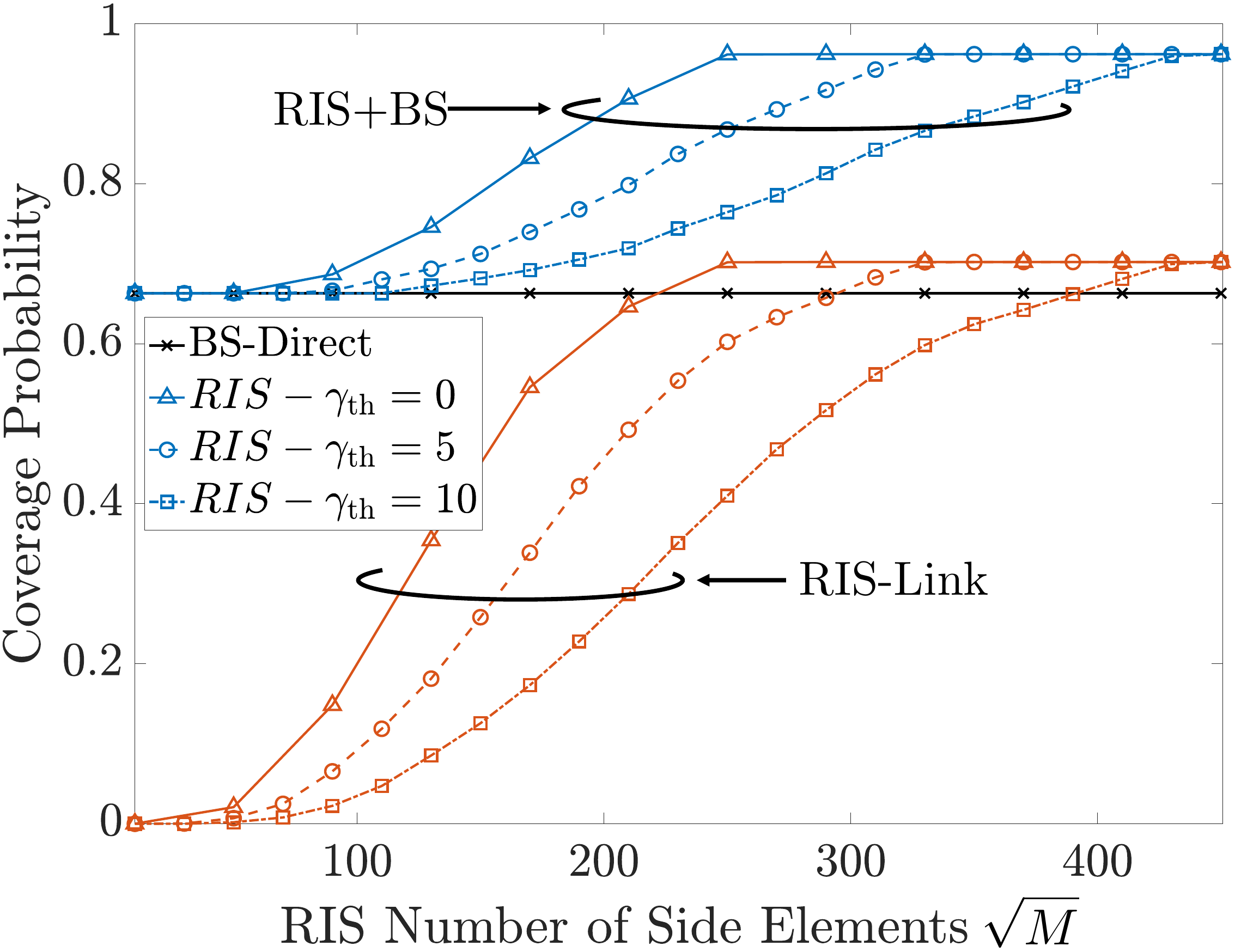}\label{fig:SC2_RIS}} \hspace{0.5cm}
\subfloat[][\small NCR-aided, urban scenario 2]{\includegraphics[width=0.42\textwidth]{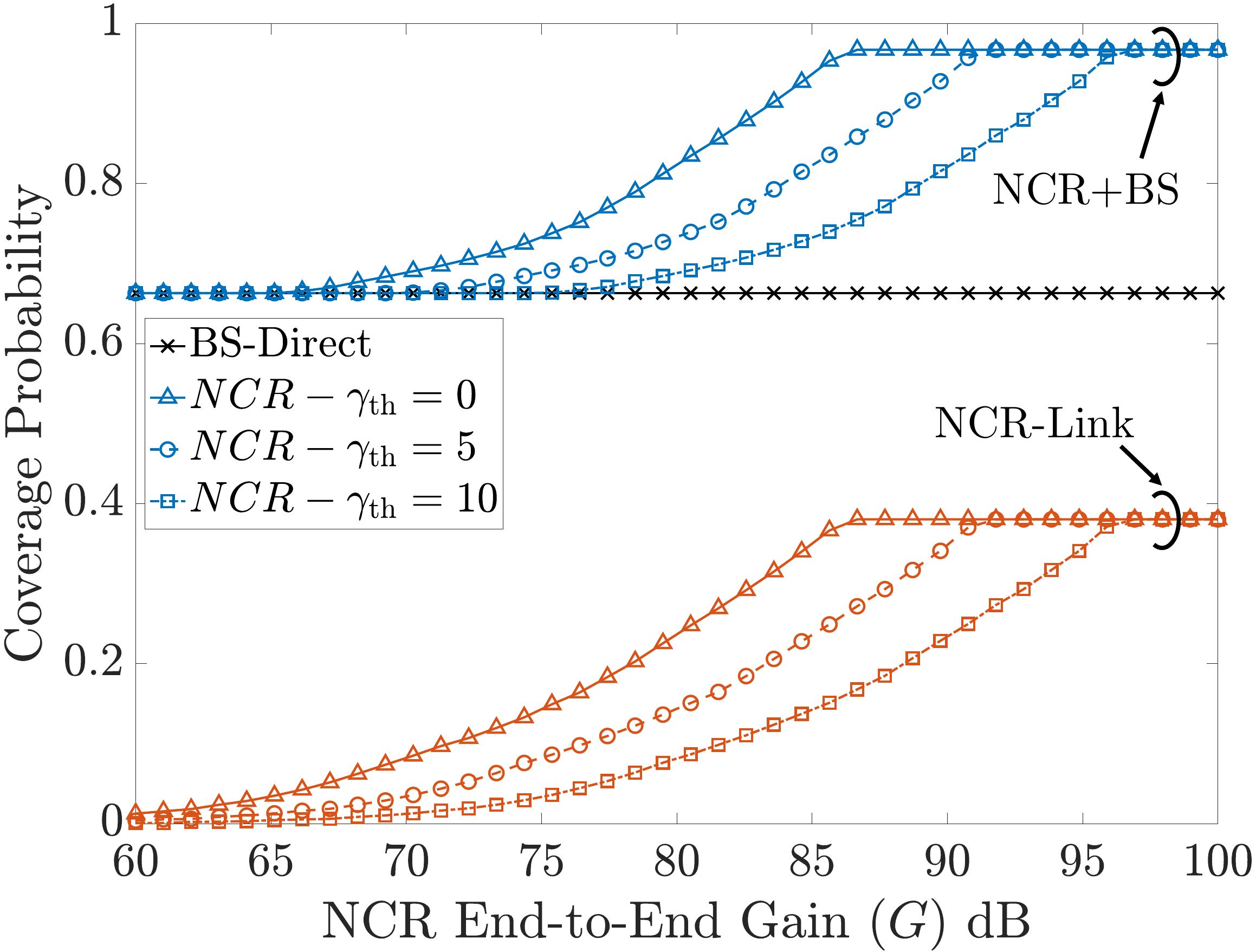}\label{fig:SC2_SR}}
        \caption{\small Coverage probability of RIS and NCR-aided communication for urban scenario 1 and urban scenario 2}
        \label{fig:mean and std of nets}
    \end{figure*}

\section{Numerical Results}\label{sect:results}
%\subsection{SNR and coverage}
\subsection{Coverage Probability}\label{sec:Coverage}
We herein compare the RIS and AF NCR by evaluating the long-term coverage probability. The latter is defined starting from the SNR in the two cases, computed as  
\begin{align}
 \gamma_{ncr} & =  \frac{\mathrm{tr}\left(\mathbf{W}^{\mathrm{H}}(\mathbf{H}_{d}+\mathbf{H}_{ncr})\mathbf{F}\mathbf{F}^{\mathrm{H}}(\mathbf{H}_{d}+\mathbf{H}_{ncr})^{\mathrm{H}}\mathbf{W}\right)}{N_r\sigma_n^2 +  \mathrm{tr}(\mathbf{H}_z\mathbf{H}_z^\mathrm{H})\sigma_z^2},\label{eq:SNR_SR} \\
 \gamma_{ris} & =  \frac{\mathrm{tr}\left(\mathbf{W}^{\mathrm{H}}(\mathbf{H}_d+\mathbf{H}_{ris})\mathbf{F}\mathbf{F}^{\mathrm{H}}(\mathbf{H}_d+\mathbf{H}_{ris})^{\mathrm{H}}\mathbf{W}\right)}{N_r\sigma_n^2}.\label{eq:SNR_RIS}
\end{align}
Precoder $\mathbf{F}$ ad combiner $\mathbf{W}$ are calculated assuming the instantaneous perfect channel knowledge at both Tx and Rx, thus via singular value decomposition (SVD) as $\mathbf{H}=\mathbf{U}\boldsymbol{\Lambda}\mathbf{V}^\mathrm{H}$, where $\mathbf{H}$ is either $\mathbf{H}_{d}+\mathbf{H}_{ncr}$ for NCR or $\mathbf{H}_d+\mathbf{H}_{ris}$ for RIS. Thus, $\mathbf{F}=\mathbf{V}_{1:N_s} \boldsymbol{\Xi}^{\frac{1}{2}}$ and $\mathbf{W} = \mathbf{U}_{1:N_s}$, where $\boldsymbol{\Xi}$ is the diagonal matrix of the allocated powers designed by water filling based on singular values $\boldsymbol{\Lambda}$ and we select the first $N_s$ columns of $\mathbf{U}$ and $\mathbf{V}$. In case of NCR, the analog Rx beamformer  $\mathbf{w}_p$ is computed based on the NCR and BS positions.% $\mathbf{p}_{R}$ and $\mathbf{p}_{BS}$ respectively. 
The Tx beamformer $\mathbf{f}_p$ is again obtained by perfect knowledge of the channel $\mathbf{H}_o$ in \eqref{eq:channel_NCR}. When only two links between Tx and Rx are available, namely direct and via the NCR/RIS relay, the BS uses one the one with the best SNR. In this latter case, we can define the SNRs of the direct and relayed links separately, namely $\gamma_{d}$ and $\gamma_{r}$ by plugging vector precoders and combiners $\mathbf{f}$ and $\mathbf{w}$ in \eqref{eq:SNR_SR} or \eqref{eq:SNR_RIS} based on the selected singular value on the channel. Given the position of BS, and UE, the long-term SNR via the direct path is 
\begin{align}
    &\overline{\gamma}_d = \mathrm{P}^{(d)}_B\,\gamma_d + (1-\mathrm{P}^{(d)}_B)\gamma^{0}_d,\label{eq:LT_SNR_direct}
\end{align}
where $\mathrm{P}^{(d)}_B$ is the blockage probability of the direct paths, $\gamma^{0}_d$ is the upper SNR limit in case the blockage is absent. The long-term SNR for the relayed path can be calculated similarly. Blockage probabilities are computed based on the assumption that the BS-relay link is never blocked, thus $\mathrm{P}^{d}_B$ is evaluated according to \eqref{eq:blockage_probabilty} putting $r=r_d = \|\mathbf{p}_{BS}-\mathbf{p}_{UE}\|_2$, while $\mathrm{P}^{r}_B$ by plugging $r=r_r = \|\mathbf{p}_{R}-\mathbf{p}_{UE}\|_2$. When both direct and relayed links are jointly used (NCR/RIS-aided communications) the average SNR is computed by considering all the blockage combinations on each link. A UE in $\mathbf{p}_{UE}$ with with generic long-term SNR $\overline{\gamma}$, is considered to be served if $\overline{\gamma}>\gamma_{th}$
where $\gamma_{th}$ is a threshold SNR. Let $\mathcal{P}'\subseteq \mathcal{P}$ be the subset of served (covered) UE positions out of the whole $\mathcal{P}$. The coverage probability can be calculated as
\begin{equation}
    \mathrm{P}_C = \frac{\lvert \mathcal{P}'\lvert}{\lvert\mathcal{P}\lvert}.\label{eq:PCov}
\end{equation}

\subsection{Simulations and Results}\label{subsect:simulations}

\begin{table}[!b]
\centering
\footnotesize
\caption{Simulation parameters \label{tab:CalcParams}}
\renewcommand{\arraystretch}{1.1}
\begin{tabular}{l|c|c}
\hline
\textbf{Parameter}                                                  & \textbf{Parameter}             & \textbf{Value} \\ \hline
Carrier frequency & $f_0$ & 28 GHz\\
Bandwidth                                                         & $B$                & 200  MHz \\
BS Tx power                                      & $\sigma_s^2$         & 35 dBm\\ 
BS array size                                                & $N_h \times N_v$        & $16 \times 12$\\
BS height                                                      & $h_{BS}$          & 6 m \\
NCR/RIS height                                                      & $h_{r}$           & 4 m
\\
NCR array size                                                    & $N_{p,h}\times N_{p,v}$           & $12 \times 6$ 
\\
NCR amplification gain                                                    & $\vert  g\vert^2$           & $55$ dB 
\\
NCR E2E gain                                                     & $G$           & $92$ dB 
\\

NCR noise figure                                                    & $NF_{ncr}$           & $8$ dB 
\\
NCR panel separation                                                         & $\alpha$           & 120 deg
\\   
RIS number of elements                         & $M_h \times M_v$           & $200 \times 200$ 
\\
RIS directivity parameter \cite{CIRS}                                                    & $q$           & $0.029$ 
\\ 
UE height                                                      & $h_{UE}$          & 1.5 m
\\

UE noise figure                                                    & $NF_{ue}$           & $10$ dB \\ 
Blocker height                                                      & $h_{B}$         & 1.7 m
\\ 
Blocker density                                                      & $\lambda_{B}$           & $4\times 10^{-3}$ m$^{-2}$\\
Blocker velocity                            & $V$           & 15 m/s
\\
Blockage duration                            & $\mu^{-1}$            & 5 s
\\ 

%BS panel tilt                                                         & $\theta^{BS}_{tilt}$               & 8 deg  \\ 
%NCR panel 2 tilt                                                         & $\theta^{ncr}_{tilt}$               & 5 deg\\ 
%shielding angle                                                      & $\theta_{sh}$   & 60  deg\\
\hline
\end{tabular}
\end{table}

The simulation setup consists of two portions of the city of Milan, Italy, portrayed in Fig. \ref{fig:Heatmap} and Fig. \ref{fig:Heatmap2}. A tri-sectoral BS is equipped with panels of $N_h \times N_v$ antennas each (along horizontal/azimuth and vertical/elevation directions, respectively) and serves the UEs in the potential coverage area $\mathcal{P}$ with both direct and relayed link (NCR or RIS). The NCR consists of two panels of $N_{p,h}\times N_{p,v}$ and the RIS consists of $M_h \times M_v$ elements. Each sector of the BS and NCR covers an angular range of $\pm 60$ deg in azimuth (a limitation imposed both by panel shielding against interference and steering) and $\pm 30$ deg in elevation (limited by steering). Locations of BS, NCR and RIS are shown in Fig. \ref{fig:Heatmap} and Fig. \ref{fig:Heatmap2}, while $\mathcal{P}$ comprises all the locations for which there exists a geometric path between the UE and one among BS, NCR and RIS. This means that the set of served positions is $\mathcal{P}'\subseteq \mathcal{P}$. The simulations are carried out by considering a single-antenna UE placed over the 3D position grid $\mathcal{P}$. The coverage probability of the considered area can be calculated according to \eqref{eq:PCov}. Otherwise noted, the simulation parameters are the ones in Table \ref{tab:CalcParams}. The directivity gain parameters of the patch antenna for BS and NCR are considered the same as in 3GPP recommendation \cite{3GPP}.
%We consider the simplified convex shapes of a portion of the city of Milan. The scenario is the downlink transmission with a tri-sectoral BS, where each panel of the BS is equipped with a uniform planar array (UPA) with horizontal and vertical number of antennas $N_h \times N_v$, located on the top left of the map (see Fig.\ref{fig:Dir}). The NCR (and RIS) is considered on the right side middle (See Fig.\ref{fig:SR} and Fig.\ref{fig:RIS}). The receiver is a generic omni-directional single antenna user equipment (UE), and the average coverage probability of the area is calculated by placing the UE in a subset of positions $\mathbf{p_{UE}} \in {\mathcal{P}}^{\prime}\subseteq \mathcal{P}$, where $ \mathcal{P}$ spans the grid of positions throughout the map portion, while $\mathcal{{P}}^{\prime}$ is a subset of servable positions, either directly by BS or indirectly via a relay. The coverage probability of the considered area can be calculated according to \eqref{eq:PCov}. The simulation parameters are shown in Table.\ref{tab:CalcParams}. Each panel of the BS and NCR serve only the UEs seen within the shielding azimuth $\theta_{sh}$.

Fig. \ref{fig:Heatmap} and Fig. \ref{fig:Heatmap2} show the heatmap of the SNR over $\mathcal{P}$, %i.e., all the possible positions taken by the UE for an urban scenario, 
when: \textit{(i)} only the direct BS-UE link is used (Fig. \ref{fig:Dir} and Fig.\ref{subfig:Dir2}); \textit{(ii)} only the RIS is used (Fig. \ref{fig:RIS} and Fig.\ref{subfig:RIS2}); \textit{(iii)} only the NCR is used (Fig. \ref{fig:SR} and Fig.\ref{subfig:SR2}). We assume that the BS-relay (NCR/RIS) is never blocked, while for all the other types of links, the blockage model in Section \ref{sect:blockage} is applied. As can be observed, according to the selected parameters, the area served by the direct BS-UE link experiences the highest SNR. However, the coverage is limited. By employing a RIS, the geometric coverage can be improved to basically include all the potential serving area. This is due to the reconfigurability properties of the RIS, that allow reaching the UE under any reflection angle. With a two-panel NCR only, oriented as in Fig. \ref{fig:SR}, we can serve a reduced area compared to the RIS but with a higher SNR on average. Differently from Fig. \ref{fig:Heatmap}, Fig. \ref{fig:Heatmap2} is a corridor-like urban scenario. Although in general, the RIS is more flexible for different AoD, the second scenario favours the usage of NCR, because the UEs along the corridor-like area would suffer a large path-loss with RIS and the NCR can better compensate for this due to its inherent amplification gain.

Fig. \ref{fig:SC1_RIS} and Fig. \ref{fig:SC1_SR} show the coverage probability \eqref{eq:PCov} for urban scenario 1, using a RIS and a NCR, varying the number of RIS element per-side $\sqrt{M}$ (Fig. \ref{fig:SC1_RIS}) and E2E gain $G$ (Fig. \ref{fig:SC1_SR}), for different SNR thresholds $\gamma_{th}$. The NCR E2E gain $G$ can be manipulated either by changing the power amplification gain $\lvert g\vert^{2}$ or the number of antennas at each NCR panel. We compare the case in which only RIS/NCR are used to the case in which the joint direct+RIS/NCR is employed (here referred to RIS/NCR-aided coverage). The SNR is evaluated with the method in Section \ref{sec:Coverage}. 
In scenario 1, the NCR-aided coverage probability saturates to $P_c \approx 0.8$ (Fig. \ref{fig:SC1_SR}), irrespective of $G$, because of the limited portion of UEs that NCR can serve, due to FoV limitations and minimum angular separation between panels $\alpha$. Fig. \ref{fig:SC2_RIS} and Fig. \ref{fig:SC2_SR} report the same analysis for urban scenario 2. In this case, the NCR-aided coverage probability saturates to $P_c \approx 0.96$ (Fig. \ref{fig:SC2_SR}). Increasing $\gamma_{th}$ means increasing the required NCR E2E gain $G$ or the number of RIS elements $M$, as expected. For instance, considering $P_c = 0.8$ at $\gamma_{th}=5$ dB as a target coverage probability, we need $M =200 \times 200$ RIS elements in both scenario 1 and scenario 2. The NCR requires E2E gain $G=95$ dB in the first scenario, while $G=85$ dB in the second scenario. As expected by observing Fig.\ref{fig:SR} and Fig.\ref{subfig:SR2}, corridor-like scenarios such as the considered one (scenario 2) favour the usage of NCR with respect to RIS. Therefore, given the knowledge of the environment, the choice between RIS and NCR has to be evaluated case-by-case, considering technical and economic aspects such as implementation and maintenance costs and signalling overhead, target QoS, etc. By and large, scenarios in which the UEs can be served within a limited FoV is the natural application for NCR, while wide-open areas can be served by RIS.

\section{Conclusion}\label{sect:conclusion}
This paper analyzes the coverage problem for mmW networks in an urban scenario, considering two network entities as possible coverage extenders, i.e., reconfigurable intelligent surfaces (RIS) and network-controlled repeater (NCR). RIS and NCR are being considered by the 3GPP standardization body to support beyond 5G and 6G network operations. The analysis is carried out considering a realistic scenario from digital maps, including the static and dynamic blockages, as per 3GPP recommendations. The numerical results show that both technologies can increase the coverage and limit the blockage impact on communcation reliability. The practical choice between RIS and NCR depends on many factors, including environment geometry, the angular separation between NCR panels, the number of antenna elements, deployment and maintenance costs, etc. The results suggest hybrid usage of NCR and RIS in large-scale network planning, whereas wide-open spaces are preferable for RIS, while narrow scenarios, e.g., open roads, are promising candidates for NCR deployment.

%In this paper a coverage analysis of the RIS-aided communication vs the NCR-aided communication is performed. It is shown that in addition to the capabilities of the RIS and NCR such as the number of reflecting/radiating elements and amplification gain, their intrinsic limitations also play a crucial role on their potential to increase the network coverage, because the areas that can be served better with RIS or NCR depend on: i) the angular separation of the relay panels dictated by the residual self-interference suppression measures; ii) antenna element directivity pattern and panel shielding; iii) the deterministic geometry of the environment.
%The results suggest hybrid usage of NCR and RIS in large scale network planning, supporting the result of some of the previous works in the literature. This work provides a useful tool for precise large scale network planning optimization analysis, given the digital map of each region of a city.

\section*{Acknowledgements}
The research has been carried out in the framework of the Huawei-Politecnico di Milano Joint Research Lab.

\bibliographystyle{IEEEtran}
\bibliography{biblio.bib}
\end{document}